\documentclass[sigconf]{acmart}

\AtBeginDocument{%
  }

\acmConference[ESEM '24]{18th ACM/IEEE International Symposium on Empirical Engineering and Measurement}{October 20--25, 2024}{Barcelona, Spain}

\usepackage{comment}
\usepackage{csquotes}
\usepackage{tabularx}
\usepackage{booktabs}
\usepackage[acronym]{glossaries}
\usepackage{listings}
\usepackage{cleveref}
\usepackage{tcolorbox}
\usepackage{todonotes}

\makeatletter
\newcommand\footnoteref[1]{\protected@xdef\@thefnmark{\ref{#1}}\@footnotemark}
\makeatother

\definecolor{CustomOrange}{HTML}{EC9D66}

\tcbuselibrary{theorems}
\usepackage{fontawesome}

\newtcbtheorem{definition1}{\faPencil{} Definition of Continuous Compliance}
{colback=blue!5,colframe=blue!35!black,fonttitle=\bfseries}{th}    

\newtcbtheorem{definition2}{\faPencil{} Definition of Continuous Security Compliance}
{colback=blue!5,colframe=blue!35!black,fonttitle=\bfseries}{th}   

\newtcbtheorem{summary}{\faInfo{} Summary}
{colback=blue!5,colframe=blue!35!black,fonttitle=\bfseries}{th} 

\newacronym{rg}{RG}{Research Goal}
\newacronym{dsm}{DSM}{Design Science Methodology}
\newacronym{cse}{CSE}{Continuous Software Engineering}
\newacronym{csc}{CSC}{\textit{Continuous Security Compliance}}

\begin{document}

\title{Towards Automated Continuous Security Compliance}


\author{Florian Angermeir}
\email{angermeir@fortiss.org}
\orcid{0000-0001-7903-8236}
\affiliation{%
  \institution{fortiss and Blekinge Institute of Technology}
  \city{Munich and Karlskrona}
  \country{Germany and Sweden}
}

\author{Jannik Fischbach}
\email{fischbach@fortiss.org}
\orcid{0000-0002-4361-6118}
\affiliation{%
  \institution{fortiss and Netlight Consulting GmbH}
  \city{Munich}
  \country{Germany}
}
  
\author{Fabiola Moy\'on}
\email{fabiola.moyon@siemens.com}
\orcid{0000-0003-0535-1371}
\affiliation{%
  \institution{Siemens Technology and Technical University of Munich}
  \city{Munich}
  \country{Germany}
}

\author{Daniel Mendez}
\email{daniel.mendez@bth.se}
\orcid{0000-0003-0619-6027}
\affiliation{%
  \institution{Blekinge Institute of Technology and fortiss}
  \city{Karlskrona and Munich}
  \country{Sweden and Germany}
}

\renewcommand{\shortauthors}{Angermeir et al.}

\begin{abstract}
Context: Continuous Software Engineering is increasingly adopted in highly regulated domains, raising the need for continuous compliance. Adherence to especially security regulations -- a major concern in highly regulated domains -- renders \emph{Continuous Security Compliance} of high relevance to industry and research. 

Problem: One key barrier to adopting continuous software engineering in the industry is the resource-intensive and error-prone nature of traditional manual security compliance activities. Automation promises to be advantageous. However, continuous security compliance is under-researched, precluding an effective adoption.

Contribution: We have initiated a long-term research project with our industry partner to address these issues. In this manuscript, we make three contributions: (1) We provide a precise definition of the term continuous security compliance aligning with the state-of-art, (2) elaborate a preliminary overview of challenges in the field of automated continuous security compliance through a tertiary literature study, and (3) present a research roadmap to address those challenges via automated continuous security compliance.
\end{abstract}



\maketitle

\section{Introduction}
Over the last two decades, \acrfull{cse}, entailing Agile and DevOps \cite{bosch2014}, transformed how software is developed, with the promise of creating software faster, providing value earlier, and, thus, resulting in higher project success rates. Some of the success factors are flexible adoption of requirements, early customer integration, enhanced team collaboration (e.g., between Dev, Ops, and Sec), and continuous experimentation and improvement \cite{chow2007, aldahmash2017, azad2023}. In the light of highly regulated domains such as healthcare, critical manufacturing, or transport systems, the same benefits entail challenges for providers, seemingly rendering \acrshort{cse} and highly regulated environments incompatible. Providers in highly regulated domains have to adhere to regulations like laws (e.g., GDPR \footnote{\url{https://data.europa.eu/eli/reg/2016/679/oj}}), standards (e.g., ISO 27001 \cite{iso27001}), internal policies, contracts or best practices to ensure the quality of their systems over disciplines including security, safety or privacy. Non-compliance with applicable regulations can negatively impact the quality of a system, with various risks for providers. These risks range from reputation damage over monetary loss\footnote{\url{https://web.archive.org/web/20230708131758/https://edpb.europa.eu/news/news/2023/12-billion-euro-fine-facebook-result-edpb-binding-decision_en}} to competitive disadvantage \cite{faustinodevops2022}. 
Providers try to reduce those risks already in requirements engineering where regulations provide a source for requirements to be extracted, analyzed as vague and volatile subjects (e.g., \cite{masseyidentifying2014, ayalariveragrace2018}), and implemented. To ensure compliance with regulations, companies must assess and demonstrate the effective fulfillment of identified requirements. In the context of \acrshort{cse}, requirements extraction, analysis, implementation, and demonstration (e.g., by verification) have to be performed frequently during development instead of following big-bang schedules, e.g. during releases \cite{fitzgerald2017}, resulting in \emph{Continuous Compliance}. However, manual compliance activities are too resource-intensive to keep up with the ever-changing nature and high development velocity of \acrshort{cse}. Already in 2014, Fitzgerald and Stol acknowledged \emph{Continuous Compliance} as a major challenge for highly regulated domains in adopting \acrshort{cse} \cite{fitzgerald2014}. Since then, more and more providers of software-intensive systems in highly regulated domains have adopted \acrshort{cse}, exacerbating the challenges and pressing need to overcome them. 

\textbf{Problem Statement.} Automation promises to be advantageous. However, the multi-faceted field of \emph{Continuous Compliance} is still vastly under-researched. Even though aspects such as automated compliance checking gained attention over recent years \cite{aberkaneexploring2021} with promising approaches such as \cite{hayrapetian2018, hamdani2021}, the need for more extensive and structured research on automation approaches is highlighted by Castellanos Ardila et al. \cite{castellanosardilacompliance2022}: \enquote{It is difficult to guarantee industrial adoption when there is nonexistent or loosely coupled tool support [...]. Thus, it is crucial to provide adequate and complete tool support for automatically \emph{[sic]} perform compliance checking} \cite{castellanosardilacompliance2022}. While automation approaches have gained attention over the last years, the compliance challenges of \acrshort{cse} in highly regulated domains and the structurization of such seem very much neglected. From a security compliance perspective, Moy\'on et al. \cite{moyon2024} offer preliminary insights. However, further research seems necessary to provide an extensive landscape of challenges in employing \acrshort{cse} while adhering to security regulations (for now referred to as \textit{Continuous Security Compliance}). Exacerbating the issue, to this day, little research has been performed on the requirements and their systematization for automation solutions in continuous compliance, even less \textit{Continuous Security Compliance}. Instead, researchers and practitioners too often follow a solution-oriented approach \cite{mendez2012} where they focus on solutions while leaving requirements underspecified and poorly understood. To our knowledge, Kellogg et al. \cite{kelloggcontinuous2020}, and Santilli et al. \cite{santilli2023} are those that stand out in literature by describing five and 13 general requirements for \textit{Continuous Security Compliance} (covering automation aspects) - in a structured way from an industrial perspective.

\textbf{Objectives.}\label{sec:researchgoals} With our research, we operate at the intersection between requirements engineering, security, and continuous software engineering. We aim to contribute to introducing automation in \textit{Continuous Security Compliance} in a problem-driven manner. To this end, we formulate the following research goals (RG):
\begin{enumerate}
    \item[\acrshort{rg}1] understand the challenges of adhering to security compliance regulations in continuous software engineering projects
    \item[\acrshort{rg}2] understand  requirements and constraints for automation in \textit{Continuous Security Compliance} as suggested in literature and practice
    \item[\acrshort{rg}3] analyze the potential and limitations of automation as treatment towards the identified challenges and its implications for manual involvement
    \item[\acrshort{rg}4] develop and evaluate -- along the previously determined potential and limitations -- sensible solutions containing automation in industrial settings to treat respective challenges
\end{enumerate}

\textbf{Contributions.} In this manuscript, we define the terms \emph{Continuous Compliance} and \emph{Continuous Security Compliance} as required for our research context. We provide the first insights into ongoing research through a tertiary study, in which we investigate and validate industrial challenges reported by practitioners. We conclude by presenting our research roadmap towards achieving our objectives and bringing together requirements engineering, security, and continuous software engineering. One hope we associate with this contribution is to foster a critical reflection and discussion on contemporary challenges in automated \textit{Continuous Security Compliance} and engage the research community to join our research endeavor.

\textbf{Data Availability.} The protocol of our literature study and the analyzed and processed data is shared in our online material \cite{onlinedocumentation}.
\section{State of Research}
Based on the current state of research, we lay the foundation for future research by defining the term \acrfull{csc} in \Cref{subsec:csc} and contributing to \acrshort{rg}1 through an analysis of relevant challenges reported in the literature in \Cref{subsec:challenges}.

\subsection{\acrfull{csc}}\label{subsec:csc}
The field of \acrshort{csc} has gained more interest over the last years. To effectively operate in this field, researchers need a definition they can identify with and refer to. While the phrase \acrlong{csc} is used in some resources (\cite{abed2016, daennert2019, rompicharla2020, alromaih2022}), to our knowledge, it has only been defined explicitly once by Moy\'on et al. as \enquote{combining CC \textit{[Continuous Compliance]} and CS \textit{[Continuous Security]} through the holistic view of standardisation that spans across people, processes, and technology. Regulatory requirements are utilised to derive security activities and integrate security into a process while making it standards-compliant} \cite{moyondefinition2020}. This definition references two concepts, \emph{continuous compliance} and \emph{continuous security}, defined by Fitzgerald and Stol in the seminal framework for continuous software engineering \textit{Continuous} * \cite{fitzgerald2014}. While the existing \acrlong{csc} definition coined the field as part of the \acrshort{cse} research agenda, it lacks preciseness concerning terminology in the compliance field (e.g. reference to regulatory requirements and compliance to standards), and has limitations on the applicability for practitioners (e.g., security compliance solely mentioned as process security compliance, assuming product security compliance is a consequence of process compliance). An updated definition is required to improve understandability based on a broader analysis of the current state-of-art, beyond \textit{Continuous} *, and, therefore, extend the concept's applicability.

We aim for a \acrlong{csc} definition that fits contemporary research while considering practice observations gathered during the last +5 years of our research on \acrshort{csc}, together with our industrial partners. 
To that end, we conducted an ad-hoc literature review (as defined by \cite{ralph2022}) on the terminology related to \textit{continuous compliance}. We reviewed 12 resources mentioning or defining continuous compliance, including theses and grey literature (\cite{fitzgerald2014, kelloggcontinuous2020, santilli2023, li2020gdpr, phipps2020, moscher2017, chari2013, shahrokni2022, abrahams2018, steffens2018, ibm2013, rysbekov2022}). We extracted relevant parts from each of the manuscripts, providing insights into the concepts covered in definitions and the actual definitions (see \cite{onlinedocumentation} for details). Most resources framed continuous compliance in light of their contribution. This confirms the observations by Santilli et al. \cite{santilli2023}, who independently analyzed 13 resources, of which three overlap with ours. Of the resources we reviewed, only two provided contribution-independent definitions. One is the original definition by Fitzgerald and Stol \cite{fitzgerald2014} from 2014, and the other was recently introduced by Santilli et al. \cite{santilli2023}. Santilli et al. also scanned the field for a proper definition and provided their version. They define continuous compliance as \enquote{the application of approaches and techniques to guarantee compliance with standards throughout a system's lifecycle. Continuous compliance activities span the various development phases [...]}. Their definition covers the most relevant aspects we could distill from the other resources as well as from our own experience (1) Continuous execution of compliance activities (e.g. assessment, update of artifacts), (2) adherence to regulatory requirements, and (3) compliance over the entire development life-cycle. However, none of the newer definitions explicitly oriented on \acrlong{cse} and its principles and goals, which is the fourth aspect we add to our definitions. In our experience, the goal of continuous compliance is as manyfold as the goals of \acrshort{cse} ranging from the enablement of compliance in fast-paced development, over enablement of organizations to respond quickly to changing compliance requirements, to continuous improvement of the compliance state.

Given these four aspects, we define the term continuous compliance ourselves and base the definition of \acrlong{csc} on it. The goal of the definitions is to not only provide a contribution-independent, yet concise updated definition but also to respect the background under which it was originally introduced: continuous software engineering. We propose to define \textit{Continuous Compliance} as a set of practices to ensure compliance targets adherence to requirements derived from relevant regulatory sources, integrated holistically into the product development life-cycle, following continuous software engineering principles and goals.

The benefits of the updated continuous compliance definition are its alignment with current state-of-the-art practices, its shortness, agnosticism regarding the regulatory source and product development methodologies, as well as the explicity orientation on continuous software engineering principles and goals. The definition covers requirements from various regulatory sources such as standards, laws, policies, or contracts and is not restricted to a certain domain (e.g. security, safety) or compliance target (e.g. software development process). While the definition ensures that the original background (\acrshort{cse}) is preserved, it does not further stipulate the product development methodology. Additionally, given that large parts of regulated domains haven't and might never fully transition to \acrshort{cse}, we do not prescribe the employed compliance process or its execution frequency. This flexibility enables practitioners to relate to \acrlong{csc} already in the early stages of the continuous improvement journey in their organization (e.g. following the stairway to heaven model of \cite{bosch2014}). The main downside of the flexibility in our definition is that it does not provide direct guidance toward implementing continuous compliance (e.g. compliance process execution frequency). However, this flaw also applies to existing definitions, and concrete guidance requires further in-depth research. With security as one application domain for continuous compliance, the before discussed advantages and disadvantages also apply to the \acrshort{csc} definition. For \acrshort{csc} the compliance target is usually products or processes and the requirements to comply with are derived from regulatory sources that are relevant to security and the respective organization.

\begin{definition2*}{}
We define \textit{\acrlong{csc} as a set of practices to ensure product and process adherence to requirements derived from relevant security regulatory sources, integrated holistically into the product development life-cycle, following continuous software engineering principles and goals.}
\end{definition2*}
\subsection{Challenges in Continuous Security Compliance}\label{subsec:challenges}

Large parts of security compliance research focus on traditional software development. This state does not reflect the ever-growing adoption of \acrfull{cse} methodologies in the industry. Consequently, effective research is difficult without structured overviews of the challenges experienced in practice when following CSE principles while pertaining to security regulations. This state is exacerbated by the difficulties of cooperating with industry on such delicate matters and publishing the findings. One instance of such research is Moy\'on et al., who present 15 challenges experienced in three different companies \cite{moyon2024}.

Contributing to \acrshort{rg}1, we research the pertinence of challenges on \acrshort{csc} in literature, validating and extending the challenges of Moy\'on et al. \cite{moyon2024}. We performed a tertiary literature study on three major scientific databases: IEEE Xplore, ACM Digital Library, and Scopus. We followed Kitchenham \cite{kitchenhamprocedures2004} for the planning and execution of the literature study. The literature review protocol and other relevant artifacts for traceability are available in our online material \cite{onlinedocumentation}. The search initially yielded 110 hits, one of which was a duplicate. Manual filtering based on the abstract and the full text left three relevant manuscripts (\cite{naegele2023, moyonchallenges18, kumar2020}).
Following Wohlin's guidelines \cite{wohlin2014}, we performed forward and backward snowballing following the references in each of the manuscripts and employing the citation feature of Google Scholar, yielding one more publication \cite{ouselati2015} (referenced by \cite{moyonchallenges18}).

The identified manuscripts investigated 15 security challenges in large-scale agile software development \cite{naegele2023}, 23 challenges while adopting DevSecOps \cite{kumar2020}, four security compliance challenges in agile software development \cite{moyonchallenges18}, and 20 challenges while developing secure software using agile under the consideration of security compliance. In total, we extracted 62 challenges from those manuscripts.

\textbf{Validation:}
Nine of the 15 challenges reported by Moy\'on et al. \cite{moyon2024} (C1, C3, C4, C6, C7, C8, C9, C13, C14) could be found in literature. Leaving six challenges unvalidated. Those challenges are not necessarily irrelevant, as further research might validate them. The challenges of \cite{moyon2024} and the number of challenges in the literature that validated them are listed in \Cref{tab:extendedchallenges} (denoted as C1-C15). The pertinency of challenges C6, C7, C8, and C14 in literature (see \Cref{tab:extendedchallenges}) highlights and aligns with Moy\'on et al.'s perspective on the need for research about automatic continuous security compliance.

\textbf{Extension:}
The remaining 47 challenges from the literature were used to extend the challenges given by Moy\'on et al. \cite{moyon2024}. To this end, we performed a thematic analysis, analyzing and aggregating similar challenges across varying granularities into 12 common new challenges (NC). The online material \cite{onlinedocumentation} details the analysis and aggregation steps, and \Cref{tab:extendedchallenges} depicts all new challenges with the number of supporting challenges from literature (denoted as NC1-NC12). For illustration purposes, we briefly describe one such new challenge. \textit{NC4: Security compliance evidence generation and documentation too resource consuming}. NC4 is based on three challenges that are thematically connected: 
\begin{itemize}
    \item \enquote{Security compliance requires documented evidence which increases delivery time} \cite{moyonchallenges18}
    \item \enquote{Security assessment favors detailed documentation} \cite{ouselati2015}
    \item \enquote{Teams need extra effort [...] to involve security experts and assessors to evaluate accuracy of documentation} \cite{moyonchallenges18}
\end{itemize}

As becomes apparent, security compliance evidence is the common subject here, with either the generation of evidence or the documentation of evidence being the activities that imply high resource demand (e.g., an increased delivery time or extra effort). Three challenges in literature could not be aggregated: \enquote{Customer readiness for applying frequent releases to production setup} \cite{kumar2020}, \enquote{Lacking guidance on the implementation of scaling agile frameworks} \cite{naegele2023}, and \enquote{Overcome security expert shortage at scale} \cite{naegele2023}.

\textbf{Usage of Challenges \& Preliminary Evaluation:}
Of the new challenges, the following seem to be relevant for automated \acrshort{csc}: NC1, NC2, NC4, NC5, NC7, and NC8. These challenges are part of the basis for our research roadmap detailed in \Cref{sec:researchroadmap}.

As a preliminary evaluation, we interviewed a security expert with more than 10 years of security experience in the industry to evaluate the understandability of the challenges and get their perspective on the most relevant challenges for automated \acrshort{csc}. 
To that end, we took a snapshot of the existing relevant challenges at that time (C6, C7, C8, C14, NC1, NC2, NC4).
We first asked for understandability of the new challenges and, afterward, which were the most relevant to them. All new challenges were perceived as easily understandable, and no clarification was necessary during the interview. The security expert perceived C6, NC1, and NC2 as the most pressing challenges in the subset providing indicators for the focus of future research (see \Cref{sec:researchroadmap}).

\begin{table}
    \footnotesize
    \centering
    \caption{\textbf{Extended List of Validated Industry Challenges} \\ C = Industry Challenges by Moy\'on et al. \cite{moyon2024}. \\ NC = New Challenges from Literature Review}
    \label{tab:extendedchallenges}
    \begin{tabularx}{\linewidth}{c X c}
        \toprule
        \multicolumn{2}{c}{CHALLENGE} & \shortstack[c]{NO. OF RELATED \\ CHALLENGES \\ IN LITERATURE} \\
        \midrule
        \multicolumn{3}{c}{\textbf{Category 1: Security in Continuous Development}} \\
        \midrule
        C1 & Perform threat modeling and consistently increment / adapt it throughout sprints & 1 \\
        C2 & Include security activities into continuous deployment & 0 \\
        C3 & Involve security aspects into continuous experimentation or prototyping & 1 \\
        NC1 & Lacking integration of security activities in development method & 6 \\
        NC2 & Changes to requirement, design or implementation break system security requirements & 3 \\
        \midrule
        \multicolumn{3}{c}{\textbf{Category 2: Security in the Value Stream}} \\
        \midrule
        C4 & Prioritization of security requirements vs. system functionalities & 3 \\
        C5 & Make security architecture visible in backlog and documentation & 0 \\
        C6 & Get security activities into the early feedback principle of DevOps & 1\\
        NC3 & Security compliance is treated as conflicting objective & 3 \\
        \midrule
        \multicolumn{3}{c}{\textbf{Category 3: Security Implementation Efficiency}} \\
        \midrule
        C7 & Balance efforts for security: improving process compliance vs. improving product quality & 2 \\
        C8 & Involve security activities with minimal burden of lead time & 2 \\
        NC4 & Security compliance evidence generation and documentation too time consuming & 3 \\
        NC5 & Security compliance too resource-intensive & 4 \\
        \midrule
        \multicolumn{3}{c}{\textbf{Category 4: Security Knowledge}} \\
        \midrule
        C9 & Enable security knowledge and ownership in engineering teams & 2 \\
        C10 & Congregate experts to interchange examples about DevOps and security & 0 \\
        NC6 & Lacking security awareness of non-engineering stakeholders & 2 \\
        \midrule
        \multicolumn{3}{c}{\textbf{Category 5: Security into CI/CD pipelines}} \\
        \midrule
        C11 & Effective hardening of development, test and deployment environments & 0 \\
        C12 & Achieve efficient handling of security tool findings and involve into regular issues handling process & 0 \\
        C13 & Selection of security tools to run in the CI/CD pipeline & 2 \\
        C14 & Match security compliance requirements with working pipelines & 1 \\
        C15 & Ensure protection of CI/CD pipelines & 0 \\
        NC7 & Limitation of security tests & 2 \\
        NC8 & Complexity in provisioning of tools and environments & 5 \\
        \midrule
        \multicolumn{3}{c}{\textbf{(New) Category 6: Security Strategy Success}} \\
        \midrule
        NC9 & Insufficient leadership on security & 5 \\
        NC10 & Collaboration and team dependencies as breakpoints & 5 \\
        NC11 & Organization of roles and responsibilities & 3 \\
        NC12 & Transformation to DevSecOps & 3 \\
        \bottomrule
    \end{tabularx}
\end{table}

\begin{summary*}{}
Nine challenges delineated by Moy\'on et al. \cite{moyon2024} were validated in literature: C1, C3, C4, C6 - C9, C13, C14. The list of challenges was extended by 12 challenges.
Challenges relevant for future automated \acrshort{csc} research: C6, C7, C8, C14, NC1, NC2, NC4, NC5, NC7, NC8.
\end{summary*}

\begin{figure}
    \centering
    \includegraphics[width=0.8\columnwidth]{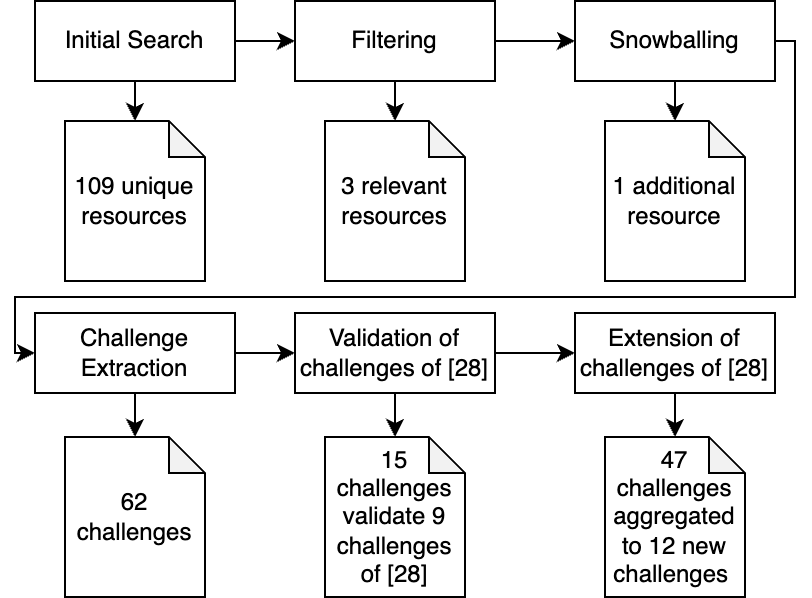}
    \caption{Literature Review Process}
    \label{fig:literature_review_process}
\end{figure}
\begin{figure*}
    \centering
    \includegraphics[width=\linewidth]{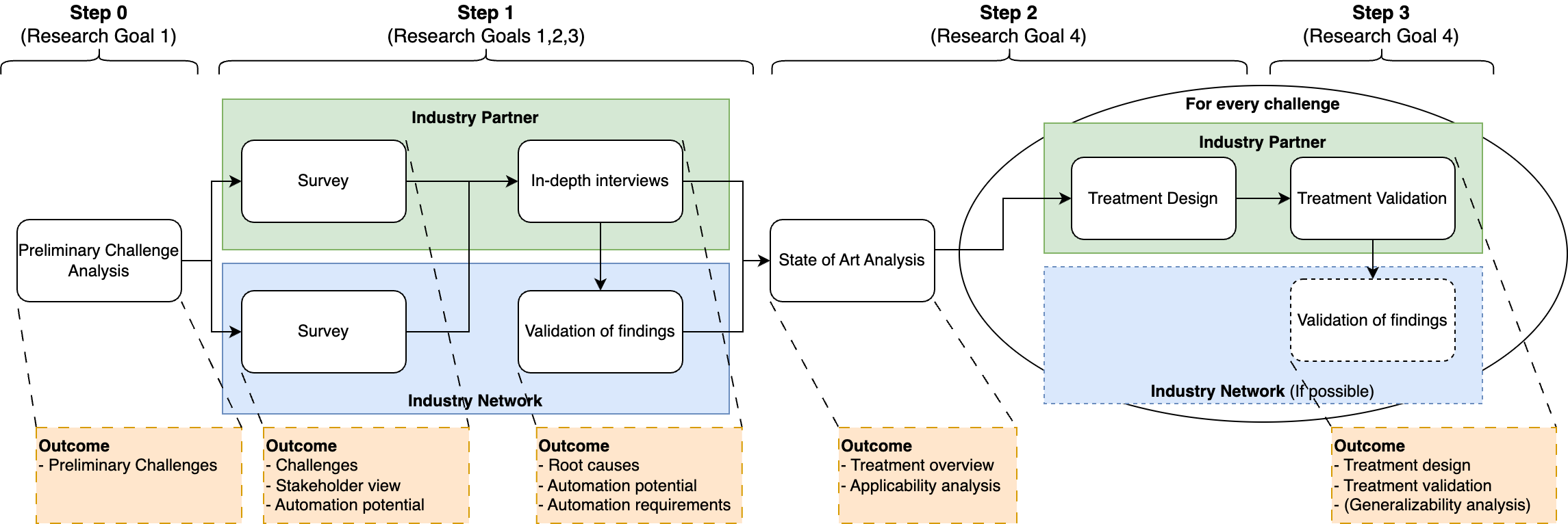}
    \caption{Summary of the Research Roadmap}
    \label{fig:researchroadmap}
\end{figure*}

\section{Research Roadmap}\label{sec:researchroadmap}
This research has four goals, as outlined in \Cref{sec:researchgoals}. In the following, we will give an overview of the research environment and outline our research vision to address the research goals. \Cref{fig:researchroadmap} summarizes the endeavor, highlighting research activities, the environments in which they are carried out, and research outcomes.

\subsection{Research Environment}
Our research is performed with an industry partner, a large enterprise operating in highly regulated domains, aligning and building on top of their strong research foundation of the last 6 years (selected examples: \cite{moyonchallenges18, moyondefinition2020, voggenreiter2022, voggenreiter2024}) in the area of compliance to security standards applicable to the development and operation of software-intensive industrial automation and control systems. To ensure generalizability, we extend and validate our findings wherever possible with industrial companies partnering with Blekinge Institute of Technology as part of Software Engineering ReThought\footnote{\label{note1}https://rethought.se/}, a large-scale academia-industry collaboration network entailing diverse industrial and academic backgrounds.

\subsection{Research Vision}\label{subsec:researchvision}
We structure the research according to the \acrfull{dsm} \cite{wieringadesign2014}, offering an iterative framework towards the problem investigation, treatment design, and treatment validation. DSM further enables the technology transfer to real-world application, closing the engineering cycle described by Wieringa. Since the \acrshort{dsm} describes a cyclic approach, we will perform steps 2 and 3 iteratively, adapting to changing requirements, challenges, and technologies. 

\textbf{Step 0. Preliminary Challenges} lays the foundation for the problem investigation of the \acrshort{dsm}. We analyzed and reported the challenges in \acrshort{csc} (\acrshort{rg}1) in \Cref{subsec:challenges}. These preliminary challenges are used in the next step of our research.

\textbf{Step 1. Challenges, Root Causes, Requirements} contributes to the problem investigation and addresses parts of the treatment design of the \acrshort{dsm}. We analyze the challenges in \acrshort{csc} and their root causes (\acrshort{rg}1), the potential of these challenges for automation (\acrshort{rg}3), and the requirements for automation in continuous security compliance (\acrshort{rg}2). 

In the first stage, we perform a survey to gather a broad perspective on the challenges and their automation potential from different stakeholders. This activity is performed internally at our industry partner and, to scale up, in parallel in the entire network of the large-scale academia-industry collaboration \textit{Software Engineering ReThought}\footnoteref{note1} as well as with selected external partners. Based on the survey findings, in the second stage, we design and perform in-depth interviews with \acrshort{cse} stakeholders targeting the root causes of the challenges, their automation potential, and respective limitations and requirements for their automated treatments. In all of these cases, we will thoroughly monitor literature (e.g. \cite{constanteintegration2020} reports on automation limitations, \cite{santilli2023} lists relevant requirements).

\textbf{Step 2. State of the Art Analysis and Solution Design} to progress in the treatment design of the \acrshort{dsm}. Together with our industry partner, we decide on a subset of challenges to focus on. To scan for available solutions to address those challenges, we will perform a technology screening through a systematic literature review. Given the existing gap between research and industry, we will additionally review the potentials, limitations, and use cases of available solutions on the market (such as Drata\footnote{https://drata.com/}, Vanta\footnote{https://www.vanta.com/}, Scrut\footnote{https://www.scrut.io/}) in the hope of better understanding and ultimately bridging this gap. We will continuously experiment to test the applicability of chosen automation approaches to the practical context. This shall provide a basis for our treatment design in which we will adopt existing approaches or design new ones (\acrshort{rg}4). We then continue with the treatment validation in step 3.

\textbf{Step 3. Treatment Validation} to close the \acrshort{dsm} cycle. We validate existing and newly developed treatments for (1) the satisfaction of the requirements identified before, (2) the impact on \acrlong{csc}, and (3) the value added by the treatments to \acrshort{cse} projects. For that purpose, we not only integrate the treatments into the processes of \acrshort{cse} projects at our industry partner but also measure their impact on \acrshort{cse} principles and consequently on their development process. Further, we analyze with security experts (e.g. auditors) whether the treatments' outputs improve \acrlong{csc}, e.g. provide valid evidence for compliance. This way, highly regulated organizations can integrate our proposed treatments into their projects while ensuring the treatment's acceptance by auditors. 

\begin{summary*}{}
We perform this research in an industrial setting iteratively in three stages: (1) research of challenges in practice, (2) analysis and design of treatments for these challenges, and (3) validation of the treatment designs.
\end{summary*}
\section{Threats to Validity}
We describe the threats to validity structured by the three contributions reported in this manuscript. First, the definition of continuous security compliance is based on an ad-hoc literature review. To put it in Ralph's and Baltes' way: The limitations of ad hoc review are the \enquote{unsystematic nature} that might be subject to sampling bias and hinders replication \cite{ralph2022}.
However, given the large content-wise overlap in-between literature and also with our insights gained from the industrial environment, we are confident that the threat is a minor one.

Second, the extension and validation of challenges in continuous security compliance. As with any systematic (tertiary) literature review, the exclusion of relevant resources is a major threat. We experimented with and improved the search strategy in four iterations before we reviewed the title and abstract of each manuscript. We only dismissed manuscripts if we were certain about their irrelevance to us. Otherwise, we read the entire manuscript. After our analysis and synthesis, the first and the third authors discussed the challenges stemming from the extension to reduce experimenters' bias. In addition, we interviewed a security expert to ensure the validated and extended challenges were suitable for further research in practice.

Finally, the research roadmap describes our long-term research projects spanning multiple years which might be subject to change. As with any research roadmap, it is exposed to various (subjective) factors influencing its creation. Although our roadmap is rooted in industrially relevant challenges and corroborated with literature results, strengthening our confidence, we cannot guarantee its relevance. As an additional means to mitigate related threats, we intentionally share our roadmap with this paper to receive feedback and support from the research community. 

\section{Conclusion}
The growing adoption of \acrfull{cse} in highly regulated domains raises the need for \acrlong{csc}. Traditional compliance activities are slow, labor-intensive, and costly, rendering them unsuited for the fast-paced development and release of contemporary software development. Automation could contribute to transitioning security compliance from traditional, rigid, and slow to continuous, flexible, and fast-paced. However, to this day, automated \acrlong{csc} is under-researched. This state forces companies operating in highly regulated domains to trade off between fast-paced development following \acrshort{cse} principles and reduced security risks by adhering to security regulations. We contributed to closing this gap by 1. providing a precise yet agnostic up-to-date definition of \acrlong{csc} reflecting its four main components: continuous execution of compliance activities, adherence to relevant security regulatory sources, compliance over the entire development life-cycle, and its continuous software engineering background. 2. we offered a preliminary extension and validation of industrial challenges at the intersection of requirements engineering, security compliance, and continuous software engineering. We also distilled six new challenges for future research of automated \acrlong{csc}. 3. we outlined a research roadmap entailing the industrial research environment, research activities, and the outcomes required to treat the identified challenges successfully. With this manuscript, we hope to foster a critical reflection of the contemporary challenges and potential treatments of automated \acrlong{csc} and engage the research community to join our research endeavor. 

\bibliographystyle{ACM-Reference-Format}
\bibliography{main}


\begin{thebibliography}{44}


\ifx \showCODEN    \undefined \def \showCODEN     #1{\unskip}     \fi
\ifx \showDOI      \undefined \def \showDOI       #1{#1}\fi
\ifx \showISBNx    \undefined \def \showISBNx     #1{\unskip}     \fi
\ifx \showISBNxiii \undefined \def \showISBNxiii  #1{\unskip}     \fi
\ifx \showISSN     \undefined \def \showISSN      #1{\unskip}     \fi
\ifx \showLCCN     \undefined \def \showLCCN      #1{\unskip}     \fi
\ifx \shownote     \undefined \def \shownote      #1{#1}          \fi
\ifx \showarticletitle \undefined \def \showarticletitle #1{#1}   \fi
\ifx \showURL      \undefined \def \showURL       {\relax}        \fi
\providecommand\bibfield[2]{#2}
\providecommand\bibinfo[2]{#2}
\providecommand\natexlab[1]{#1}
\providecommand\showeprint[2][]{arXiv:#2}

\bibitem[Abed et~al\mbox{.}(2016)]%
        {abed2016}
\bibfield{author}{\bibinfo{person}{Javad Abed}, \bibinfo{person}{Gurpreet Dhillon}, {and} \bibinfo{person}{Sevgi Ozkan}.} \bibinfo{year}{2016}\natexlab{}.
\newblock \showarticletitle{Investigating Continuous Security Compliance Behavior: Insights from Information Systems Continuance Model}. In \bibinfo{booktitle}{\emph{AMCIS '16}}. \bibinfo{numpages}{10}~pages.
\newblock


\bibitem[Aberkane et~al\mbox{.}(2021)]%
        {aberkaneexploring2021}
\bibfield{author}{\bibinfo{person}{Abdel-Jaouad Aberkane}, \bibinfo{person}{Geert Poels}, {and} \bibinfo{person}{Seppe~Vanden Broucke}.} \bibinfo{year}{2021}\natexlab{}.
\newblock \showarticletitle{Exploring {Automated} {GDPR}-{Compliance} in {Requirements} {Engineering}: {A} {Systematic} {Mapping} {Study}}.
\newblock \bibinfo{journal}{\emph{IEEE Access}}  \bibinfo{volume}{9} (\bibinfo{date}{5} \bibinfo{year}{2021}), \bibinfo{pages}{66542--66559}.
\newblock


\bibitem[Abrahams and Langerman(2018)]%
        {abrahams2018}
\bibfield{author}{\bibinfo{person}{Muhammad~Zaid Abrahams} {and} \bibinfo{person}{Josef~J Langerman}.} \bibinfo{year}{2018}\natexlab{}.
\newblock \showarticletitle{Compliance at Velocity within a DevOps Environment}. In \bibinfo{booktitle}{\emph{ICDIM '18}}. \bibinfo{pages}{94--101}.
\newblock


\bibitem[Aldahmash et~al\mbox{.}(2017)]%
        {aldahmash2017}
\bibfield{author}{\bibinfo{person}{Abdullah Aldahmash}, \bibinfo{person}{Andy~M. Gravell}, {and} \bibinfo{person}{Yvonne Howard}.} \bibinfo{year}{2017}\natexlab{}.
\newblock \bibinfo{booktitle}{\emph{Systems, Software and Services Process Improvement}}.
\newblock Chapter A Review on the Critical Success Factors of Agile Software Development, \bibinfo{pages}{504--512}.
\newblock


\bibitem[Alromaih et~al\mbox{.}(2022)]%
        {alromaih2022}
\bibfield{author}{\bibinfo{person}{Arwa Alromaih}, \bibinfo{person}{Yasser Ismail}, {and} \bibinfo{person}{Wael Elmedany}.} \bibinfo{year}{2022}\natexlab{}.
\newblock \showarticletitle{Continuous compliance to ensure strong cybersecurity posture within digital transformation in smart cities}. In \bibinfo{booktitle}{\emph{SCS '22}}. \bibinfo{pages}{464--479}.
\newblock


\bibitem[Angermeir et~al\mbox{.}(2024)]%
        {onlinedocumentation}
\bibfield{author}{\bibinfo{person}{Florian Angermeir}, \bibinfo{person}{Jannik Fischbach}, \bibinfo{person}{Fabiola Moy{\'o}n}, {and} \bibinfo{person}{Daniel Mendez}.} \bibinfo{year}{2024}\natexlab{}.
\newblock \bibinfo{title}{Towards Automated Continuous Security Compliance}.
\newblock
\newblock
\urldef\tempurl%
\url{https://doi.org/10.6084/m9.figshare.25199225.v1}
\showURL{%
\tempurl}


\bibitem[Ayala-Rivera and Pasquale(2018)]%
        {ayalariveragrace2018}
\bibfield{author}{\bibinfo{person}{Vanessa Ayala-Rivera} {and} \bibinfo{person}{Liliana Pasquale}.} \bibinfo{year}{2018}\natexlab{}.
\newblock \showarticletitle{The {Grace} {Period} {Has} {Ended}: {An} {Approach} to {Operationalize} {GDPR} {Requirements}}. In \bibinfo{booktitle}{\emph{RE'18}}. \bibinfo{pages}{136--146}.
\newblock


\bibitem[Azad and Hyrynsalmi(2023)]%
        {azad2023}
\bibfield{author}{\bibinfo{person}{Nasreen Azad} {and} \bibinfo{person}{Sami Hyrynsalmi}.} \bibinfo{year}{2023}\natexlab{}.
\newblock \showarticletitle{DevOps critical success factors — A systematic literature review}.
\newblock \bibinfo{journal}{\emph{Information and Software Technology}}  \bibinfo{volume}{157} (\bibinfo{date}{5} \bibinfo{year}{2023}), \bibinfo{numpages}{14}~pages.
\newblock


\bibitem[Bosch(2014)]%
        {bosch2014}
\bibfield{editor}{\bibinfo{person}{Jan Bosch}} (Ed.). \bibinfo{year}{2014}\natexlab{}.
\newblock \bibinfo{booktitle}{\emph{Continuous Software Engineering}}.
\newblock \bibinfo{publisher}{Springer}.
\newblock


\bibitem[Castellanos~Ardila et~al\mbox{.}(2022)]%
        {castellanosardilacompliance2022}
\bibfield{author}{\bibinfo{person}{Julieth~Patricia Castellanos~Ardila}, \bibinfo{person}{Barbara Gallina}, {and} \bibinfo{person}{Faiz Ul~Muram}.} \bibinfo{year}{2022}\natexlab{}.
\newblock \showarticletitle{Compliance checking of software processes: {A} systematic literature review}.
\newblock \bibinfo{journal}{\emph{Journal of Software: Evolution and Process}} \bibinfo{volume}{34}, \bibinfo{number}{5} (\bibinfo{date}{5} \bibinfo{year}{2022}), \bibinfo{numpages}{36}~pages.
\newblock


\bibitem[Chari et~al\mbox{.}(2013)]%
        {chari2013}
\bibfield{author}{\bibinfo{person}{Suresh Chari}, \bibinfo{person}{Ian Molloy}, \bibinfo{person}{Youngja Park}, {and} \bibinfo{person}{Wilfried Teiken}.} \bibinfo{year}{2013}\natexlab{}.
\newblock \showarticletitle{Ensuring continuous compliance through reconciling policy with usage}. In \bibinfo{booktitle}{\emph{SACMAT '13}}. \bibinfo{pages}{49–--60}.
\newblock


\bibitem[Chow and Cao(2008)]%
        {chow2007}
\bibfield{author}{\bibinfo{person}{Tsun Chow} {and} \bibinfo{person}{Dac-Buu Cao}.} \bibinfo{year}{2008}\natexlab{}.
\newblock \showarticletitle{A survey study of critical success factors in agile software projects}.
\newblock \bibinfo{journal}{\emph{Journal of Systems and Software}} \bibinfo{volume}{81}, \bibinfo{number}{6} (\bibinfo{date}{6} \bibinfo{year}{2008}), \bibinfo{pages}{961--971}.
\newblock


\bibitem[Constante et~al\mbox{.}(2020)]%
        {constanteintegration2020}
\bibfield{author}{\bibinfo{person}{Fabiola~Moy\'on Constante}, \bibinfo{person}{Rafael Soares}, \bibinfo{person}{Maria Pinto-Albuquerque}, \bibinfo{person}{Daniel Mendez}, {and} \bibinfo{person}{Kristian Beckers}.} \bibinfo{year}{2020}\natexlab{}.
\newblock \showarticletitle{Integration of Security Standards in {DevOps} Pipelines: An Industry Case Study}.
\newblock In \bibinfo{booktitle}{\emph{PROFES '20}}. \bibinfo{pages}{434--452}.
\newblock


\bibitem[D{\"a}nnart et~al\mbox{.}(2019)]%
        {daennert2019}
\bibfield{author}{\bibinfo{person}{Sebastian D{\"a}nnart}, \bibinfo{person}{Fabiola~Moy{\'o}n Constante}, {and} \bibinfo{person}{Kristian Beckers}.} \bibinfo{year}{2019}\natexlab{}.
\newblock \showarticletitle{An Assessment Model for Continuous Security Compliance in Large Scale Agile Environments}. In \bibinfo{booktitle}{\emph{Advanced Information Systems Engineering}}. \bibinfo{pages}{529--544}.
\newblock


\bibitem[Faustino et~al\mbox{.}(2022)]%
        {faustinodevops2022}
\bibfield{author}{\bibinfo{person}{João Faustino}, \bibinfo{person}{Daniel Adriano}, \bibinfo{person}{Ricardo Amaro}, \bibinfo{person}{Rubén Pereira}, {and} \bibinfo{person}{Miguel~Mira da Silva}.} \bibinfo{year}{2022}\natexlab{}.
\newblock \showarticletitle{{DevOps} benefits: {A} systematic literature review}.
\newblock \bibinfo{journal}{\emph{Software: Practice and Experience}} \bibinfo{volume}{52}, \bibinfo{number}{9} (\bibinfo{date}{9} \bibinfo{year}{2022}), \bibinfo{pages}{1905--1926}.
\newblock


\bibitem[Fitzgerald and Stol(2014)]%
        {fitzgerald2014}
\bibfield{author}{\bibinfo{person}{Brian Fitzgerald} {and} \bibinfo{person}{Klaas-Jan Stol}.} \bibinfo{year}{2014}\natexlab{}.
\newblock \showarticletitle{Continuous Software Engineering and beyond: Trends and Challenges}. In \bibinfo{booktitle}{\emph{RCoSE '14}}. \bibinfo{pages}{1–9}.
\newblock


\bibitem[Fitzgerald and Stol(2017)]%
        {fitzgerald2017}
\bibfield{author}{\bibinfo{person}{Brian Fitzgerald} {and} \bibinfo{person}{Klaas-Jan Stol}.} \bibinfo{year}{2017}\natexlab{}.
\newblock \showarticletitle{Continuous software engineering: A roadmap and agenda}.
\newblock \bibinfo{journal}{\emph{Journal of Systems and Software}}  \bibinfo{volume}{123} (\bibinfo{date}{1} \bibinfo{year}{2017}), \bibinfo{pages}{176--189}.
\newblock


\bibitem[Hamdani et~al\mbox{.}(2021)]%
        {hamdani2021}
\bibfield{author}{\bibinfo{person}{Rajaa~El Hamdani}, \bibinfo{person}{Majd Mustapha}, \bibinfo{person}{David~Restrepo Amariles}, \bibinfo{person}{Aurore Troussel}, \bibinfo{person}{Sébastien Meeùs}, {and} \bibinfo{person}{Katsiaryna Krasnashchok}.} \bibinfo{year}{2021}\natexlab{}.
\newblock \showarticletitle{A combined rule-based and machine learning approach for automated {GDPR} compliance checking}. In \bibinfo{booktitle}{\emph{ICAIL '21}}. \bibinfo{pages}{40--49}.
\newblock


\bibitem[Hayrapetian and Raje(2018)]%
        {hayrapetian2018}
\bibfield{author}{\bibinfo{person}{Allenoush Hayrapetian} {and} \bibinfo{person}{Rajeev Raje}.} \bibinfo{year}{2018}\natexlab{}.
\newblock \showarticletitle{Empirically {Analyzing} and {Evaluating} {Security} {Features} in {Software} {Requirements}}. In \bibinfo{booktitle}{\emph{ISEC '18}}. \bibinfo{pages}{1--11}.
\newblock


\bibitem[IBM(2013)]%
        {ibm2013}
\bibfield{author}{\bibinfo{person}{IBM}.} \bibinfo{year}{2013}\natexlab{}.
\newblock \bibinfo{title}{Maintaining continuous compliance—a new best-practice approach}.
\newblock
\newblock
\urldef\tempurl%
\url{https://docs.media.bitpipe.com/io_11x/io_115656/item_894327/Maintaining%20continuous%20compliance.pdf}
\showURL{%
\tempurl}


\bibitem[{International Standards Organization}(2018)]%
        {iso27001}
\bibfield{author}{\bibinfo{person}{{International Standards Organization}}.} \bibinfo{year}{2018}\natexlab{}.
\newblock \bibinfo{booktitle}{\emph{Information technology - Security techniques - Information security management systems}}.
\newblock \bibinfo{type}{ISO Standard} 27001.
\newblock


\bibitem[Kellogg et~al\mbox{.}(2020)]%
        {kelloggcontinuous2020}
\bibfield{author}{\bibinfo{person}{Martin Kellogg}, \bibinfo{person}{Martin Schäf}, \bibinfo{person}{Serdar Tasiran}, {and} \bibinfo{person}{Michael~D. Ernst}.} \bibinfo{year}{2020}\natexlab{}.
\newblock \showarticletitle{Continuous Compliance}. In \bibinfo{booktitle}{\emph{ASE '20}}. \bibinfo{pages}{511--523}.
\newblock


\bibitem[Kitchenham(2004)]%
        {kitchenhamprocedures2004}
\bibfield{author}{\bibinfo{person}{Barbara Kitchenham}.} \bibinfo{year}{2004}\natexlab{}.
\newblock \bibinfo{booktitle}{\emph{Procedures for {Performing} {Systematic} {Reviews}}}.
\newblock \bibinfo{type}{{T}echnical {R}eport}. \bibinfo{address}{Keele, UK and Eveleigh, Australia}.
\newblock


\bibitem[Kumar and Goyal(2020)]%
        {kumar2020}
\bibfield{author}{\bibinfo{person}{Rakesh Kumar} {and} \bibinfo{person}{Rinkaj Goyal}.} \bibinfo{year}{2020}\natexlab{}.
\newblock \showarticletitle{Modeling continuous security: A conceptual model for automated DevSecOps using open-source software over cloud (ADOC)}.
\newblock \bibinfo{journal}{\emph{Computer Security}} \bibinfo{volume}{97}, \bibinfo{number}{C} (\bibinfo{date}{10} \bibinfo{year}{2020}), \bibinfo{numpages}{28}~pages.
\newblock


\bibitem[Li et~al\mbox{.}(2020)]%
        {li2020gdpr}
\bibfield{author}{\bibinfo{person}{Ze~Shi Li}, \bibinfo{person}{Colin Werner}, \bibinfo{person}{Neil Ernst}, {and} \bibinfo{person}{Daniela Damian}.} \bibinfo{year}{2020}\natexlab{}.
\newblock \showarticletitle{GDPR Compliance in the Context of Continuous Integration}.
\newblock  (\bibinfo{year}{2020}).
\newblock
\urldef\tempurl%
\url{https://doi.org/10.48550/arXiv.2002.06830}
\showDOI{\tempurl}
\showeprint{arXiv:2002.06830v1}


\bibitem[Massey et~al\mbox{.}(2014)]%
        {masseyidentifying2014}
\bibfield{author}{\bibinfo{person}{Aaron~K. Massey}, \bibinfo{person}{Richard~L. Rutledge}, \bibinfo{person}{Annie~I. Anton}, {and} \bibinfo{person}{Peter~P. Swire}.} \bibinfo{year}{2014}\natexlab{}.
\newblock \showarticletitle{Identifying and classifying ambiguity for regulatory requirements}. In \bibinfo{booktitle}{\emph{RE'14}}. \bibinfo{pages}{83--92}.
\newblock


\bibitem[Moscher(2017)]%
        {moscher2017}
\bibfield{author}{\bibinfo{person}{Marco Moscher}.} \bibinfo{year}{2017}\natexlab{}.
\newblock \emph{\bibinfo{title}{Continuous Compliance Testing}}.
\newblock \bibinfo{thesistype}{Master's\ thesis}.
\newblock


\bibitem[Moy\'{o}n et~al\mbox{.}(2024)]%
        {moyon2024}
\bibfield{author}{\bibinfo{person}{Fabiola Moy\'{o}n}, \bibinfo{person}{Florian Angermeir}, {and} \bibinfo{person}{Daniel Mendez}.} \bibinfo{year}{2024}\natexlab{}.
\newblock \showarticletitle{Industrial Challenges in Secure Continuous Development}. In \bibinfo{booktitle}{\emph{ICSE '24}}. \bibinfo{numpages}{3}~pages.
\newblock


\bibitem[Moy\'on et~al\mbox{.}(2018)]%
        {moyonchallenges18}
\bibfield{author}{\bibinfo{person}{Fabiola Moy\'on}, \bibinfo{person}{Kristian Beckers}, \bibinfo{person}{Sebastian Klepper}, \bibinfo{person}{Philipp Lachberger}, {and} \bibinfo{person}{Bernd Bruegge}.} \bibinfo{year}{2018}\natexlab{}.
\newblock \showarticletitle{Towards continuous security compliance in agile software development at scale}. In \bibinfo{booktitle}{\emph{RCoSE '18}}. \bibinfo{pages}{31–34}.
\newblock


\bibitem[Moy\'{o}n et~al\mbox{.}(2020)]%
        {moyondefinition2020}
\bibfield{author}{\bibinfo{person}{Fabiola Moy\'{o}n}, \bibinfo{person}{Daniel M{\'e}ndez}, \bibinfo{person}{Kristian Beckers}, {and} \bibinfo{person}{Sebastian Klepper}.} \bibinfo{year}{2020}\natexlab{}.
\newblock \showarticletitle{How to Integrate Security Compliance Requirements with Agile Software Engineering at Scale?}. In \bibinfo{booktitle}{\emph{PROFES '20}}. \bibinfo{pages}{69--87}.
\newblock


\bibitem[{Méndez Fernández} et~al\mbox{.}(2012)]%
        {mendez2012}
\bibfield{author}{\bibinfo{person}{Daniel {Méndez Fernández}}, \bibinfo{person}{Stefan Wagner}, \bibinfo{person}{Klaus Lochmann}, \bibinfo{person}{Andrea Baumann}, {and} \bibinfo{person}{Holger {de Carne}}.} \bibinfo{year}{2012}\natexlab{}.
\newblock \showarticletitle{Field study on requirements engineering: Investigation of artefacts, project parameters, and execution strategies}.
\newblock \bibinfo{journal}{\emph{Information and Software Technology}} \bibinfo{volume}{54}, \bibinfo{number}{2} (\bibinfo{date}{2} \bibinfo{year}{2012}), \bibinfo{pages}{162--178}.
\newblock


\bibitem[Nägele et~al\mbox{.}(2023)]%
        {naegele2023}
\bibfield{author}{\bibinfo{person}{Sebastian Nägele}, \bibinfo{person}{Natalie Schenk}, {and} \bibinfo{person}{Florian Matthes}.} \bibinfo{year}{2023}\natexlab{}.
\newblock \showarticletitle{The Current State of Security Governance and Compliance in Large-Scale Agile Development: A Systematic Literature Review and Interview Study}. In \bibinfo{booktitle}{\emph{CBI '23}}. \bibinfo{pages}{1--10}.
\newblock


\bibitem[Oueslati et~al\mbox{.}(2015)]%
        {ouselati2015}
\bibfield{author}{\bibinfo{person}{Hela Oueslati}, \bibinfo{person}{Mohammad~Masudur Rahman}, {and} \bibinfo{person}{Lotfi~ben Othmane}.} \bibinfo{year}{2015}\natexlab{}.
\newblock \showarticletitle{Literature Review of the Challenges of Developing Secure Software Using the Agile Approach}. In \bibinfo{booktitle}{\emph{ARES '15}}. \bibinfo{pages}{540–547}.
\newblock


\bibitem[Phipps and Zacchiroli(2020)]%
        {phipps2020}
\bibfield{author}{\bibinfo{person}{Simon Phipps} {and} \bibinfo{person}{Stefano Zacchiroli}.} \bibinfo{year}{2020}\natexlab{}.
\newblock \showarticletitle{Continuous Open Source License Compliance}.
\newblock \bibinfo{journal}{\emph{Computer}} \bibinfo{volume}{53}, \bibinfo{number}{12} (\bibinfo{date}{12} \bibinfo{year}{2020}), \bibinfo{pages}{115--119}.
\newblock


\bibitem[Ralph and Baltes(2022)]%
        {ralph2022}
\bibfield{author}{\bibinfo{person}{Paul Ralph} {and} \bibinfo{person}{Sebastian Baltes}.} \bibinfo{year}{2022}\natexlab{}.
\newblock \showarticletitle{Paving the way for mature secondary research: the seven types of literature review}. In \bibinfo{booktitle}{\emph{ESEC/FSE'22}}. \bibinfo{numpages}{5}~pages.
\newblock


\bibitem[Rompicharla and P.~V(2020)]%
        {rompicharla2020}
\bibfield{author}{\bibinfo{person}{Rajesh Rompicharla} {and} \bibinfo{person}{Bhaskar~Reddy P.~V}.} \bibinfo{year}{2020}\natexlab{}.
\newblock \showarticletitle{Continuous Compliance model for Hybrid Multi-Cloud through Self-Service Orchestrator}. In \bibinfo{booktitle}{\emph{ICSTCEE '20}}. \bibinfo{pages}{589--593}.
\newblock


\bibitem[Rysbekov(2022)]%
        {rysbekov2022}
\bibfield{author}{\bibinfo{person}{Arstanaly Rysbekov}.} \bibinfo{year}{2022}\natexlab{}.
\newblock \emph{\bibinfo{title}{Continuous Compliance: DevOps Approach to Compliance And Change Management}}.
\newblock \bibinfo{thesistype}{Master's\ thesis}.
\newblock


\bibitem[Santilli et~al\mbox{.}(2023)]%
        {santilli2023}
\bibfield{author}{\bibinfo{person}{Tiziano Santilli}, \bibinfo{person}{Patrizio Pelliccione}, \bibinfo{person}{Rebekka Wohlrab}, {and} \bibinfo{person}{Ali Shahrokni}.} \bibinfo{year}{2023}\natexlab{}.
\newblock \showarticletitle{What is Continuous Compliance?}
\newblock \bibinfo{journal}{\emph{IEEE Software}} (\bibinfo{date}{12} \bibinfo{year}{2023}), \bibinfo{pages}{1--10}.
\newblock


\bibitem[Shahrokni and Pelliccione(2022)]%
        {shahrokni2022}
\bibfield{author}{\bibinfo{person}{Ali Shahrokni} {and} \bibinfo{person}{Patrizio Pelliccione}.} \bibinfo{year}{2022}\natexlab{}.
\newblock \showarticletitle{Significance of Continuous Compliance in Automotive}. In \bibinfo{booktitle}{\emph{EASE '22}}. \bibinfo{pages}{272--–273}.
\newblock


\bibitem[Steffens et~al\mbox{.}(2018)]%
        {steffens2018}
\bibfield{author}{\bibinfo{person}{Andreas Steffens}, \bibinfo{person}{Horst Lichter}, {and} \bibinfo{person}{Marco Moscher}.} \bibinfo{year}{2018}\natexlab{}.
\newblock \showarticletitle{Towards Data-Driven Continuous Compliance Testing}. In \bibinfo{booktitle}{\emph{SE '18}}. \bibinfo{pages}{78--84}.
\newblock


\bibitem[Voggenreiter et~al\mbox{.}(2022)]%
        {voggenreiter2024}
\bibfield{author}{\bibinfo{person}{Markus Voggenreiter}, \bibinfo{person}{Florian Angermeir}, \bibinfo{person}{Fabiola Moy\'on}, \bibinfo{person}{Ulrich Sch\"{o}pp}, {and} \bibinfo{person}{Pierre Bonvin}.} \bibinfo{year}{2022}\natexlab{}.
\newblock \showarticletitle{Automated Security Findings Management: A Case Study in Industrial DevOps}. In \bibinfo{booktitle}{\emph{ICSE-SEIP '22}}. \bibinfo{numpages}{11}~pages.
\newblock


\bibitem[Voggenreiter and Sch\"{o}pp(2022)]%
        {voggenreiter2022}
\bibfield{author}{\bibinfo{person}{Markus Voggenreiter} {and} \bibinfo{person}{Ulrich Sch\"{o}pp}.} \bibinfo{year}{2022}\natexlab{}.
\newblock \showarticletitle{Using a semantic knowledge base to improve the management of security reports in industrial DevOps projects}. In \bibinfo{booktitle}{\emph{ICSE-SEIP '22}}. \bibinfo{pages}{309–310}.
\newblock


\bibitem[Wieringa(2014)]%
        {wieringadesign2014}
\bibfield{author}{\bibinfo{person}{Roel~J. Wieringa}.} \bibinfo{year}{2014}\natexlab{}.
\newblock \bibinfo{booktitle}{\emph{Design Science Methodology for Information Systems and Software Engineering}}.
\newblock


\bibitem[Wohlin(2014)]%
        {wohlin2014}
\bibfield{author}{\bibinfo{person}{Claes Wohlin}.} \bibinfo{year}{2014}\natexlab{}.
\newblock \showarticletitle{Guidelines for snowballing in systematic literature studies and a replication in software engineering}. In \bibinfo{booktitle}{\emph{EASE '14}}. Article \bibinfo{articleno}{38}, \bibinfo{numpages}{10}~pages.
\newblock


\end{thebibliography}

\appendix

\end{document}